\documentclass[twocolumn,showpacs,prb]{revtex4}
\usepackage{graphicx}
\usepackage{dcolumn}
\usepackage{amsmath}
\begin{document}
\title{On the phase of oscillatory microwave photoresistance and zero-resistance states}
\author{M. A. Zudov}
\affiliation{Department of Physics, University of Utah, Salt Lake City, Utah 84112}
\received{19 June 2003}
\begin{abstract}
We present phenomenological analysis of the period and the phase of oscillatory microwave photoresistance (OMP)\cite{zudovprb} and zero-resistance states (ZRS)\cite{mani, zudovzrs} recently observed in 2D electron systems.
The results demonstrate that as OMP evolves into ZRS with increasing magnetic field, the absolute value of the phase becomes progressively smaller, decreasing roughly as $1/B$.
Virtually eliminating a phase-shift and resulting in different periodicities for the maxima and minima, such specific dependence is supported by a simple model based on oscillatory density of states.
Finally, it follows that fine structures first reported in Ref.~3 can be viewed as an experimental evidence for multi-photon processes.

\end{abstract}
\maketitle

Using innovative microwave (MW) photoconductivity spectroscopy of 2D electron systems (2DES) originally employed in experiments on oscillatory microwave photoresistance (OMP)\cite{zudovprb}, two research groups have recently reported on a series of ``zero-resistance states'' (ZRS)\cite{mani,zudovzrs} emerging from the OMP minima in ultra-high quality samples.
Manifesting a novel dissipationless regime, such states appear when the MW frequency, $\omega=2\pi f$, somewhat exceeds the cyclotron frequency, $\omega_c=eB/m$, of the 2DES ($m$ is the effective electron mass) and are characterized by an exponentially small low-temperature resistance and a classical Hall resistance.
More recently, experiments have been extended to probe {\em dc} conduc\-ti\-vity in Corbino rings of 2DES revealing ``zero-conductance states'' (ZCS),\cite{zcs} in agreement with standard {\em dc} magneto-transport tensor relation.
Discovery of ZRS has triggered a surge of theoretical interest\cite{phillips,durst,andreev,anderson,shi,koulakov,volkov,mikhailov,bergeret,dmitriev,cheremisin,dorozhkin,lei,rivera,ryzhii,lee,ryzhii2,vavilov,klesse,volkov2,ryzhii3} and has been confirmed in independent experiments.\cite{willettaps,dorozhkin}
As it was realized that even the mechanism of the original OMP lacks understanding, the first step forward was made by Durst et al\cite{durst} who related the phenomenon to radiation-induced impurity-assisted scattering (in fact, similar ideas were proposed decades ago by Ryzhii.\cite{ryzhii69,ryzhii86}).
Regardless of the microscopic nature of the OMP, it was established experimentally\cite{zudov1} that the sample mobility favors OMP amplitude at the same time reducing background resistance.
Therefore, one could intuitively expect that further improving sample quality would eventually lead to zero or even negative resistance at the OMP minima.
As the later scenario is not experimentally realized,\cite{mani,zudovzrs,zcs} Andreev et al\cite{andreev} presented strong arguments showing that a negative resistance(conductance) state, regardless of its origin, is intrinsically unstable.
This instability leads to formation of current(electric field) domains\cite{andreev,anderson,bergeret} which give rise to ZRS(ZCS). 
As there are no new experiments available to date to test these or other theories, current understanding of the phenomenon appears far from complete. 
Furthermore, while there seems to be a consensus about the period of the ZRS, the value of the phase seems to be controversial, even experimentally.\cite{mani,zudovzrs}
Since the majority of proposed OMP(ZRS) models seem to account for, and some [see, e.g. Ref.~2] even rely on a specific value of the phase, we feel that it would be useful to address the origin of this discrepancy.

In this paper we present detailed analysis of the period and the phase of the OMP/ZRS, which, we hope, complements original experimental findings.\cite{mani,zudovzrs}
The results indicate that evolution of the OMP into ZRS with increasing magnetic field is accompanied by a dramatic reduction of the phase.
We find that in the ZRS regime the phase decreases roughly as $1/B$ and such dependence virtually eliminates a ``$1/4$-cycle'' phase shift attributed to ZRS by the authors of Ref.~2.
While a specific origin of such a dependence is not clear at this point, we show that, under reasonable assumptions, it is consistent with an idea\cite{durst} that MW photoresistance roughly follows the derivative of the density of states (DoS).
Finally, we show that fine structures first reported in Ref.~3 can be viewed as an experimental evidence for multi-photon processes.

One of the puzzles surrounding experimental reports is that, despite a great deal of similarity between the data presented in Ref.~2 and that of Ref.~3, conclusions regarding the phase of the ZRS were quite different.
Mani et al\cite{mani} have found the positions of the maxima/minima ($^\pm$, respectively) of both the OMP and ZRS to be described by:
\begin{equation}
\varepsilon^{\pm}_j=j \mp 1/4
\label{1}
\end{equation}
where $\varepsilon \equiv \omega/\omega_c$ and $j$ is a positive integer.
According to Eq.~(1) maxima(minima) appear blue(red)-shifted by a ``$1/4$-cycle'' from the cyclotron resonance harmonics, $\varepsilon_j=j$, and the magnetoresistance can be viewed as a single-harmonic function with the phase, $\phi^{\pm}=\mp0.25$.

On the other hand, Zudov et al\cite{zudovzrs} have reported that {\em major} ($j\lesssim 4$) maxima can be roughly fitted to $\varepsilon^{+}\approx j$.
As far as the major minima are concerned, their positions, contrary to the maxima, are not well defined since ZRS span a wide range of the magnetic field. 
Naively one could take the ZRS center as its position but the higher temperature data\cite{zudovzrs} and apparent asymmetry of the major maxima\cite{zudovzrs} rule against such single-harmonic picture.
Therefore, Zudov et al\cite{zudovzrs} have proposed that ZRS could also be viewed as a roughly periodic (in $1/B$) sequence, with no apparent phase shift, although with a somewhat enhanced periodicity [cf., Fig.~3 in Ref.~3].
Here we summarize the observations of Ref.~3 as follows:
\begin{equation}
\varepsilon^{\pm}_j = \alpha^{\pm} j, \,\,\,\,j\lesssim 4
\label{2}
\end{equation}
where $\alpha^{\pm}$ is a constant close to unity.
Higher-order ($j\gtrsim 4$) OMP were found to conform to Eq.~(1) although such approach required a somewhat reduced value of the effective mass ($m_{\rm lo}=0.064m_0$), as opposed to $m_{\rm hi}=0.068$ obtained using Eq.~(2) for $j\lesssim 4$.\cite{zudovaps}


\begin{figure}[b]
\includegraphics{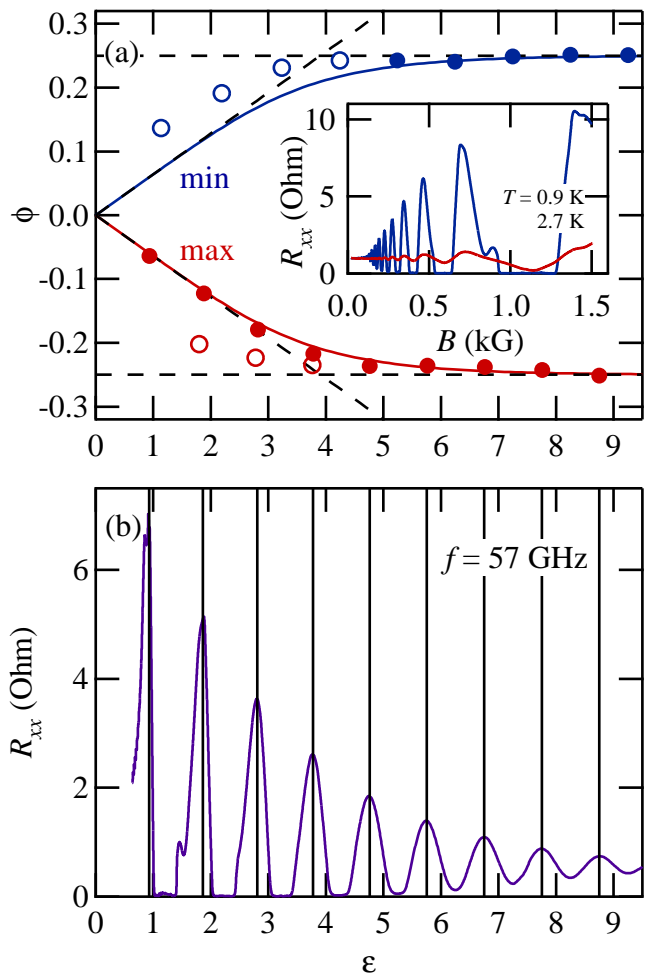}
\caption{(a) Solid(open) circles are experimental values of the phase, $\phi^{\pm}=\varepsilon^{\pm}_j - j$, extracted from the data\cite{zudovzrs} (inset) taken at $T$ = 0.9 K($T$ = 2.7 K) and $f$ = 57 GHz.
Solid lines: $\phi^{\pm}$ calculated using Eq.~(\ref{rc1}) for $m=0.064$ and $\Gamma=0.3$ K.
Dashed lines are asymptotes of Eq.~(\ref{rc1}) calculated using the same parameters.
(b) Magnetoresistance under MW illumination of $f$ = 57 GHz taken at $T$ = 0.9 K (also seen in the inset of Fig.~1(a)) plotted as a function of $\varepsilon$.
Vertical lines are calculated using Eqs.~(\ref{rc0}),(\ref{rc1}) for the maxima.
}
\label{fig1}
\end{figure}
It immediately follows that, experimentally, the boundary of applicability of Eqs.~(\ref{1}) and (\ref{2}), i.e., $j\approx 4$, seems to also separate the regimes of the OMP from that of the ZRS; being described by different equations MW photoresistance for $j\gtrsim4$ corresponds to the OMP regime, while ZRS appear at $j\lesssim 4$.
To reconcile Eqs.~(1) and (2) we propose the following expression
\begin{equation}
\varepsilon^{\pm}_j=j+\phi^{\pm},
\label{rc0}
\end{equation}
where $\phi^{\pm}$ is now allowed to vary with $\varepsilon$, approaching $\mp 0.25$ as $\varepsilon$ increases.
In what follows, we analyse the phase, $\phi^\pm_j\equiv \varepsilon^\pm_j-j$, extracted from our experimental data,\cite{zudovzrs,samples2} as a function of $\varepsilon$.\cite{effmass}

In Fig.~1(a) we present the experimental value of the phase, $\phi^{\pm}$, extracted from magnetoresistance traces\cite{zudovzrs} (see inset) taken at $T$ = 0.9 K (solid circles) and $T$ = 2.7 K (open circles) for $f=57$ GHz.
Horizontal dashed lines mark a ``$\mp$1/4-cycle'' phase-shift, which is readily observed in experiment for both  maxima and minima at $\varepsilon\gtrsim 4$.
At $\varepsilon \lesssim 4$, positions of the minima at low temperature cannot be accurately determined and we limit our discussion to the maxima positions (as discussed later in the text, minima positions are expected to follow similar dependence).
With decreasing $\varepsilon$ we observe a dramatic reduction of $|\phi^+|$, roughly linear with $\varepsilon$, i.e., $\phi^+\approx-\beta^+\varepsilon$, with $\beta^+\approx 6.4\times10^{-2}\ll 1$.
We immediately note that, in agreement with earlier conclusions,\cite{zudovzrs} such a dependence does not produce any phase-shift but affects periodicity [cf., Eq.~(2)].
Indeed, substituting this result into Eq.~(\ref{rc0}) one obtains $\varepsilon^{+}=j/(1+\beta^+)\approx (1-\beta^+)j$ which is just Eq.~(2).
We can now relate the phenomenological parameter $\beta^+$ to the difference between the effective masses  extracted earlier\cite{zudovaps} from the maxima positions using Eq.~(1) for $j\gtrsim4$ ($m_{\rm lo}=0.064$) and Eq.~(2) (with $\alpha^+ =1$) for $j\lesssim4$ ($m_{\rm hi}=0.068$).
We note that completely ignoring the phase-shift results in an overestimation of the mass by approximately $\beta^+$~\%, consistent with the data of Fig.~1(a).
While the minima positions at $\varepsilon \lesssim 4$ could be accessed at elevated temperatures, e.g. $T$ = 2.7 K, the extracted phase (open circles) does not seem to follow such a simple linear dependence on $\varepsilon$. 
While the same is true for the maxima at this $T$, the origin of such enhancement of the phase remains unclear.

It is interesting to examine the extracted phase in terms of the recent theoretical proposals.
Here we chose a ``toy model'' proposed by Durst et al\cite{durst} who suggested that MW photoresistance roughly follows the derivative of the DoS taken at $E=\hbar\omega$.
The condition describing the positions of the maxima and minima in the OMP(ZRS) structure is then given by:
\begin{equation}
\left.\frac{d^2N(E)}{dE^2}\right |_{E=\hbar\omega}=0
\label{rc}
\end{equation}
Using a well-known fact that in weak magnetic field the oscillatory part of the DoS behaves as $\cos(2\pi E/\hbar\omega_c)$, Eq.~(1) is easily recovered.
It is well known, however, that with increasing magnetic field, cyclotron energy will eventually exceed the Landau level (LL) width and DoS will no longer be described by a single-harmonic function.
Due to this, in regular magnetotransport, Shubnikov-de Haas (SdH) effect evolves into a quantum Hall effect (QHE) with increasing magnetic field (or sample mobility).
Intuitively, one could think that similar magnetic-field-driven transition might be responsible for the evolution of the OMP\cite{zudovprb} into ZRS.\cite{mani,zudovzrs}
Qualitatively it is straightforward to see that as the DoS deviates from a sinusoidal form with increasing magnetic field, the phase will be {\em reduced} rendering Eq.~(1) invalid and therefore irrelevant to the ZRS regime.

For quantitative comparison with experiment we assume that LLs have Lorentzian shape characterized by a field-independent width, $\Gamma$.\cite{SCBA}
Then the DoS can be written as:
\begin{equation}
N(E)=\frac{1}{\pi^2\ell_0^2}\sum_{n}\frac{\Gamma}{(E-n\hbar\omega_c)^2+\Gamma^2},
\label{dos}
\end{equation}
where $n$ denotes the LL index and $\ell_0=\sqrt {\hbar/eB}$ is the magnetic length.
Experimentally we are constrained to the case of very high LLs ($\hbar\omega\ll E_F$, $E_F$ is the Fermi energy), therefore, the summation can be taken over infinite limits yielding analytical solution.
After introducing dimensionless units ($\varepsilon=E/\hbar\omega_c$, $\gamma=\Gamma/\hbar\omega_c$, and $n(\varepsilon)=N(E)\hbar\omega_c\pi\ell_0^2$), one obtains:
\begin{equation}
n(\varepsilon)= \left[\cos^2(\pi\varepsilon)\tanh(\pi\gamma)+\sin^2(\pi\varepsilon)\coth(\pi\gamma)\right]^{-1}
\label{dosnorm}
\end{equation}
Substituting (\ref{dosnorm}) in (\ref{rc}) (i.e., $d^2n(\varepsilon)/d\varepsilon^2\left . \right |_{\varepsilon=\omega/\omega_c}$) and solving for $\varepsilon$ one obtains Eq.~(\ref{rc0}) with the phase of the form:
\begin{equation}
\phi^{\pm}=\mp\frac{1}{2\pi}\arccos\psi
\label{rc1}
\end{equation}
where $\psi=1/2-y+\sqrt{y^2-y+9/4}$, $y=\cosh^2(\pi\gamma)$, and $\phi^{\pm}$ is the phase for the series of maxima/minima, respectively.

At lower magnetic fields ($\gamma \gg 1$), $y\gg 1$, $\psi \sim y^{-1} \ll 1$, so $\phi^{\pm}\approx \mp 1/4$ and Eqs.~(\ref{rc0}),(\ref{rc1}) reduce to Eq.~(1).
One can also arrive at the same conclusion by noticing that in this limit, as mentioned earlier, the DoS becomes a single-harmonic function of $\varepsilon$, i.e., $n(\varepsilon)= 1+2\exp(-2\gamma)\cos(2\pi\varepsilon)$.

More interesting results emerge at higher magnetic fields ($\gamma \ll 1$), when LLs become well separated.
In this limit, $y\approx 1+\pi^2\gamma^2/2$, $\psi \approx 1+2\pi^2\gamma^2/3$, which leads to $\phi^{\pm} \approx \mp\gamma/\sqrt{3}$.
One can also easily obtain the same result by considering an isolated Lorentzian line.
We immediately note that the phase is decreasing as $1/B$, in agreement with experimental data plotted in Fig.~1(a), leading to $j \propto 1/B$ [cf., Eq.~(2)]:
\begin{equation}
j=\frac{\hbar\omega \pm \Gamma/\sqrt 3}{\hbar\omega_c}.
\label{rchi}
\end{equation}

A few comments are appropriate.
First, Eq.~(\ref{rchi}) is consistent with our experimental observations\cite{zudovzrs} regarding the positions of the major maxima and minima.
Oscillation order $j$ for the maxima and minima scales linearly with $\varepsilon$, with no {\em apparent} phase, but with different prefactors [cf., $\alpha^{\pm}$ in Eq.~(2)]; it is easy to see that since $\gamma\ll 1$, $\alpha^{\pm} \approx 1$, as observed experimentally.
Second, the asymmetry of the ZRS portion of the magnetoresistance trace can now be understood, since Eq.~(\ref{rchi}) dictates oscillations to appear as closely-spaced maximum-minimum pairs, centered about integer values of $\varepsilon=j$.
Third, since experimentally OMP ($j\gtrsim 4$) and ZRS ($j\lesssim 4$) conform to different resonant conditions (Eq.~(1) and Eq.~(2), respectively) this model suggests that ZRS develop from the OMP minima as a result of a magnetic-field-driven transition taking place around $\hbar\omega_c/2\Gamma \sim 1$.
Such a transition would roughly occur when $\phi^{\pm}=\mp\gamma/\sqrt 3$ reaches its low-field limit of $\mp$1/4.
Indeed, it happens when $\hbar\omega_c/2\Gamma=\sqrt{3}/2\sim 1$.
Finally, it is interesting to mention that Eq.~(\ref{rchi}) provides a direct experimental method to probe $\Gamma$, which is not directly accessible in standard magnetotransport, both in SdH ($\gamma \ll 1$) and in QHE ($\gamma \gg 1$) regime.
We also mention that more detailed microscopic calculations within a self-consistent Born approximation\cite{vavilov} predict similar reduction of the phase with increasing magnetic field.



Using Eq.~(\ref{rchi}) we can now relate the phenomenological parameter $\beta$ to $\Gamma$, i.e., $\Gamma=\sqrt 3\hbar\omega\beta \approx$ 0.3 K ($\Gamma \ll \hbar\omega \approx 2.7 $ K, as expected).
We can also roughly estimate the number of developed ZRS, as $\hbar\omega/(2\Gamma) \approx 4$, in agreement with experiment.
Using Eqs.~(\ref{rc0}),(\ref{rc1}) we now compute the maxima positions for the whole range of $\varepsilon$ for comparison with experimental data.\cite{zudovzrs}
In Fig.~1(b) we present the results of such calculations shown by vertical lines along with experimental trace for $f=57$ GHz adopted from Ref.~3, but now replotted as a function of $\varepsilon$.
While it was shown before\cite{zudovaps} that Eq.~(1) works well only for $\varepsilon\gtrsim 4$ and Eq.~(2) for $\varepsilon\lesssim 4$, Eqs.~(\ref{rc0}),(\ref{rc1}) provide excellent agreement over the whole range of $\varepsilon$, both in OMP and ZRS regimes.
In Fig.~1(a) we now present $\phi^{\pm}$, calculated using Eq.~(\ref{rc1}) for $m=0.064$ and $\Gamma=0.3$ K (solid lines), and again observe good agreement with low-temperature experimental data.\cite{zudovzrs} 
Dashed lines crossing around $\varepsilon \approx 4$ represent asymptotes of Eq.~(\ref{rc1}), i.e., $\phi^\pm \approx \mp \gamma/\sqrt 3$ ($\gamma\ll 1$) and $\phi^\pm \approx \mp 0.25$ ($\gamma\gg1$).
Experimentally, we observe that this crossing point roughly marks a transition from OMP to ZRS [cf. Fig.~1(b)].
An enhancement of the phase at higher $T$ can now be related to the thermal broadening of LLs.

\begin{figure}[t]
\includegraphics{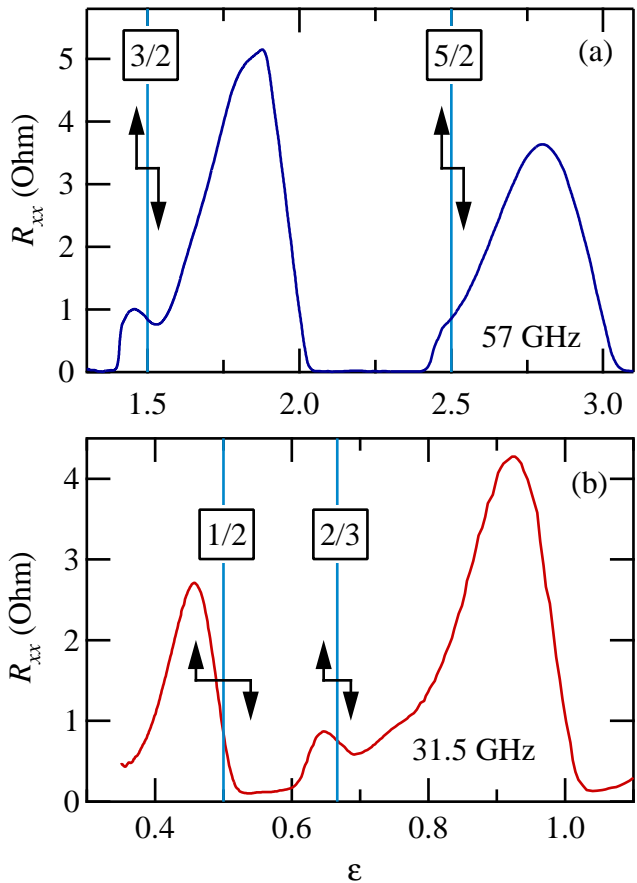}
\caption{Magnetoresistance under MW illumination of frequency (a) 57 GHz and (b) 31.5 GHz.
Vertical lines are drawn at $\varepsilon=j/m$= 3/2, 5/2 (a) and 1/2, 2/3 (b).
Vertical arrows, placed symmetrically about $j/m$ mark maximum-minimum pairs at $\varepsilon^\pm_{jm}=j/m\mp\phi_{jm}$.
}
\label{mpfig}
\end{figure}
Once we have established that OMP can be viewed as maximum-minimum pairs associated with integer $j$ and positioned around $\varepsilon=j$, we can try to generalize Eq.~(\ref{rc0}) for the processes involving multiple photons.
While it seems unlikely that such higher-order processes would be readily resolved experimentally, our data\cite{zudovzrs} suggest that such a scenario deserve close examination, especially in light of recent theoretical comments.\cite{durst, lei, maxim}
For the case of $m$-photon processes and $\gamma \ll 1$, the maximum-minimum pairs are to appear roughly symmetrically around {\em fractional} $\varepsilon$ and the expression (\ref{rc0}) is modified as follows (here we do not attempt to calculate $\phi_{jm}$):
\begin {equation}
\varepsilon^{\pm}_{jm} = \frac{j}{m} \pm \phi_{jm}
\label{mp}
\end{equation}
where $m=2,3,4...$\,.
For instance, two-photon processes ($m=2$) would reveal themselves as a series of maximum-minimum pairs close to half-integer values of $\varepsilon$, e.g. 1/2, 3/2, 5/2,...\,.
To prove the feasibility of such scenario, we present in Fig.~\ref{mpfig}(a) the magnetoresistance data\cite{zudovzrs} taken under illumination with MW radiation of $f=57$ GHz but now plotted over the narrow range of $\varepsilon$.
Maximum-minimum pair centered at $\varepsilon = 3/2$ and marked by vertical arrows is clearly observed and similar structure seems to develop around $\varepsilon = 5/2$.
While such secondary peaks appearing at $\varepsilon >1$ may be possibly explained by other mechanisms, structures emerging at $\varepsilon <1$ present stronger support for multi-photon transitions as these naturally allow to enter the region of $\varepsilon <1$.
In Fig. \ref{mpfig}(b) we show magnetoresistance data\cite{unpublished} for $f=31.5$ GHz and focus on the region of $\varepsilon<1$.
The structure centered around $\varepsilon = 1/2$ is comparable in amplitude to the primary single-photon structure around $\varepsilon=1$.
We notice that this feature is best observed at low MW frequencies\cite{zudovzrs} and quickly disappears at $f\gtrsim 40$ GHz as it gradually shifts into SdH regime.
In addition, there appears yet another maximum-minimum pair close to $\varepsilon=2/3$ which would suggest an even less-likely, three-photon process.
Based on the good agreement of the positions, we believe that secondary peaks first observed in Ref.~3 are due to multi-photon processes as described by Eq.~(\ref{mp}) for $m=2$.
The test for such a conclusion would be the systematic power-dependence experiments which are deferred for future studies.

In summary, we have studied the period and the phase of the MW photoresistance over the wide range of $\varepsilon$, covering both OMP and ZRS regimes.
As OMP evolves into ZRS with increasing magnetic field we observe a dramatic reduction of the phase, which decreases roughly as $1/B$.
Such a decrease results in different periodicities for the maxima and minima, but both exhibit no apparent phase-shift, in agreement with our earlier report.\cite{zudovzrs}
Assuming that MW photoresistance follows the derivative of the DoS,\cite{durst} ZRS and OMP can be viewed as two different experimental regimes separated by the condition $\hbar\omega_c/2\Gamma \approx 1$.
Despite obvious oversimplification, such an intuitive model seems to capture the behavior of the ZRS position/phase quite well but we do not rule out other explanations.
Finally, we identify additional structures first reported in Ref.~3 as resulting from multi-photon processes taking place around fractional values of $\varepsilon$, e.g. $\varepsilon \approx 1/2, 3/2, ...$\,.

We would like to thank R. R. Du, V. A. Apalkov, and M. G. Vavilov for reading the manuscript and valuable remarks.
This work is supported by DARPA.


\begin{thebibliography}{10}
\bibitem{samples2}
Our samples were Al$_{0.24}$Ga$_{0.76}$As/GaAs/Al$_{0.24}$Ga$_{0.76}$
  symmetrically $\delta$-doped quantum wells grown by molecular beam epitaxy,
  with a mobility $\mu=2.5 \times 10^7$ cm$^2$/Vs and an electron density
  $n_e=3.5 \times 10^{11}$ cm$^{-2}$. For additional sample parameters and
  experimental details the reader is referred to Ref.~3.
\bibitem{effmass}
Here, we chose $m=0.064m_0$ deduced from the OMP regime because it encompasses,
  experimentally, a much wider range of $\varepsilon$ as compared to the ZRS
  regime. We notice an apparent discrepancy between the deduced $m=0.064m_0$
  and the generally accepted value of $0.068m_0$ for GaAs. The difference
  ($\sim6$\%) well exceeds experimental uncertainties.
\bibitem{SCBA}
While a choice of a lineshape of the LLs (e.g., gaussian or Lorentzian) is not
  particularly important for our discussion, we notice that the assumption
  regarding the $B$-independent width of the LL contradicts to the generaly
  accepted $\sqrt{B}$-dependence obtained using a self-consitent Born
  approximation, which would lead to a $\sqrt \varepsilon$-dependence of the
  phase, $\phi$. Whether high-temperature data presented in Fig.~1(b) indicate
  such dependence remains to be clarified by future experiments. We also note
  that our choice was successfully used in SdH formalism.
\end{thebibliography}

\end{document}